\documentclass{nature}
\usepackage{graphicx}
\usepackage{placeins}
\usepackage{nature_journal_style}
\usepackage{deluxetable}
\usepackage{hyperref}

\usepackage{datetime}
\usepackage{amsmath}

\usepackage[utf8]{inputenc}

\usepackage[ngerman, num]{isodate}
\monthyearsepgerman{\,}{\,}

\usepackage{color}
\definecolor{shadecolor}{rgb}{0.9,0.9,0.9}

\title{
A common Milgromian acceleration scale in nature 
 }
\author{Pavel~Kroupa$^{1,2}$, Indranil Banik$^3$, Hosein Haghi$^{1,4}$, Akram Hasani Zonoozi$^{1,4}$, J\"org Dabringhausen$^2$, Behnam Javanmardi$^{5}$, Oliver M\"uller$^{6}$, Xufen Wu$^{7,8}$, Hongsheng Zhao$^{3,9}$}

\begin{document}

\maketitle

\begin{affiliations}
\item Helmholtz-Institut f\"ur Strahlen- und Kernphysik, University of Bonn, Nussallee 14-16, D-53115 Bonn, Germany
\item Charles University in Prague, Faculty of Mathematics and Physics, Astronomical Institute, V Hole\v{s}ovi\v{c}k\'ach 2, CZ-18000 Praha 8, Czech Republic
\item Scottish Universities' Physics Alliance, University of St Andrews, North Haugh, St Andrews, Fife, KY16 9SS, UK 
\item Institute for Advanced Studies in Basic Sciences, Physics department, Gavazang road, Zanjan, 4513766731 Zanjan, Iran
\item School of Astronomy,
Institute for Research in Fundamental Sciences (IPM) P. O. Box 19395-5531, Tehran, Iran
\item Departement Physik, Universität Basel, Klingelbergstrasse 82, CH-4056 Basel, Switzerland
\item CAS Key Laboratory for Research in Galaxies and Cosmology, Department of Astronomy, University of Science and Technology of China, Hefei, 230026, P.R. China
\item School of Astronomy and Space Science, University of Science and Technology of China, Hefei 230026, China
\item Department of Physics and Astronomy, LaserLaB, Vrije Universiteit, De Boelelaan 1081, NL-1081 HV Amsterdam, the Netherlands
\end{affiliations}

\clearpage
\textbf{
A central problem of contemporary physics is whether the law of gravity is non-Newtonian on galaxy scales. Rodrigues et al.\cite{Rodr18} argue that Milgromian gravitation\cite{FM12}, which solves the flat rotation curve problem without the need for dark matter particles, is ruled out at $>10 \sigma$ significance. To a large extent, this conclusion relies on galaxies with very uncertain distances and/or nearly edge-on orientations, where dust obscuration often becomes significant. Applying appropriate quality cuts to the data leaves only a handful of outliers to the predictions of Milgromian gravitation according to the analysis of Rodrigues et al.\cite{Rodr18}, but even these outliers can be explained with Milgromian gravitation.
}

Milgromian gravitation (MOND) is a classical theory of gravitation with a non-linear generalised Poisson equation derivable from a Lagrangian\cite{BM84}. In MOND, only the distribution of standard (luminous) matter sources the gravitational potential. To a Newtonian observer, this potential implies a dark matter halo. In regions where the acceleration is below the critical scale of Milgrom's constant 
$a_{_0} = 1.2\times 10^{-10}\,$m/s$^{2}$ $\approx 3.6\,$pc/Myr$^2$,  the equations of motion behave, possibly due to the quantum vacuum\cite{Milgrom99}, as if the system was Newtonian but with a dark matter content, elegantly solving the so-called `dark matter' problem.

Rodrigues et al.\cite{Rodr18} analyse high quality rotation curves of 100 galaxies and fit a Milgromian acceleration scale, $g_0$, to each individually. They argue that the observed dispersion in $g_0$ is inconsistent with a single value equal to $a_0$ with $> 10 \sigma$ significance, therewith falsifying MOND. 

\begin{figure} 
\includegraphics[width=12cm]{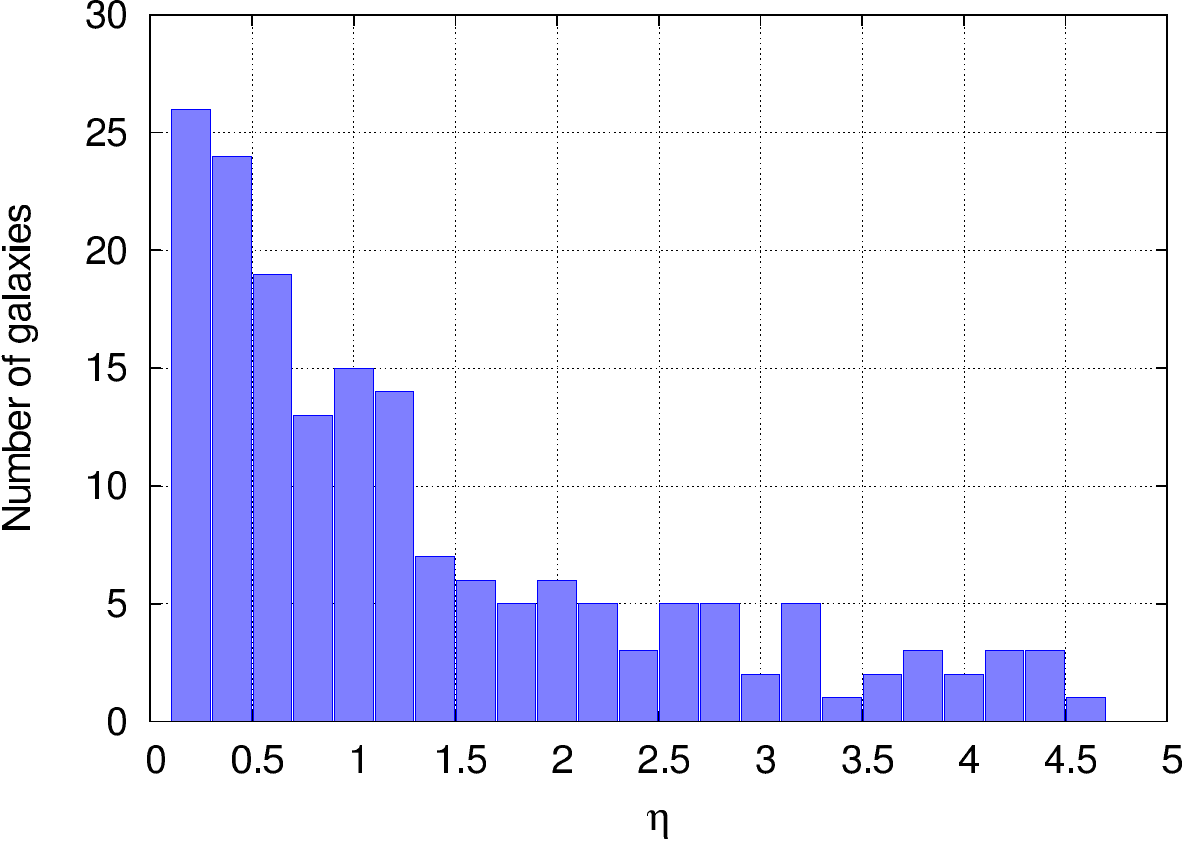} 
\caption{\bf Distribution of inclinations for all galaxies to be MONDian} {\small Assuming $g_0 = a_0$ (in contrast to Rodrigues et al.\cite{Rodr18}) and the same stellar $M/L$ for the bulges and disks as in\cite{Li+18}, for each of the 175~galaxies the inclination, $i$, is varied until $\chi^2$ of the rotation curve is minimised. The best MONDian model yields $i_{\rm best}$ for each galaxy. The $\chi^2$ remains nearly unchanged with respect to the analysis of Rodrigues et al.\cite{Rodr18} because our assumption that $i_{\rm best} \neq i_{\rm obs}$ compensates for their assumption that $g_0 \neq a_0$.  The distribution of $\eta \equiv | (i_{\rm best} - i_{\rm obs} )/ (2 \delta_{\rm obs}) |$, where $i_{\rm obs}$ and $\delta_{\rm obs}$ are the published\cite{Lelli16a} inclination and its uncertainty, respectively, is shown as the blue histogram; 98 galaxies have $\eta<1$, and 135 galaxies have $\eta<2$ and in all cases the adjustment to $i$ needed in order for each galaxy to be MONDian is within reasonable bounds on the inclination uncertainty (note that the true inclination may deviate from the published one by a few times its uncertainty).  } \label{fig:incl} \end{figure}

Importantly, Rodrigues et al.\cite{Rodr18} do not take into account inclination uncertainties.  Fig.~\ref{fig:incl} demonstrates that fixing $g_0$ to the MOND standard value of $a_0$ implies inclinations consistent with the observations within uncertainties.

\begin{figure} 
\includegraphics[width=12cm]{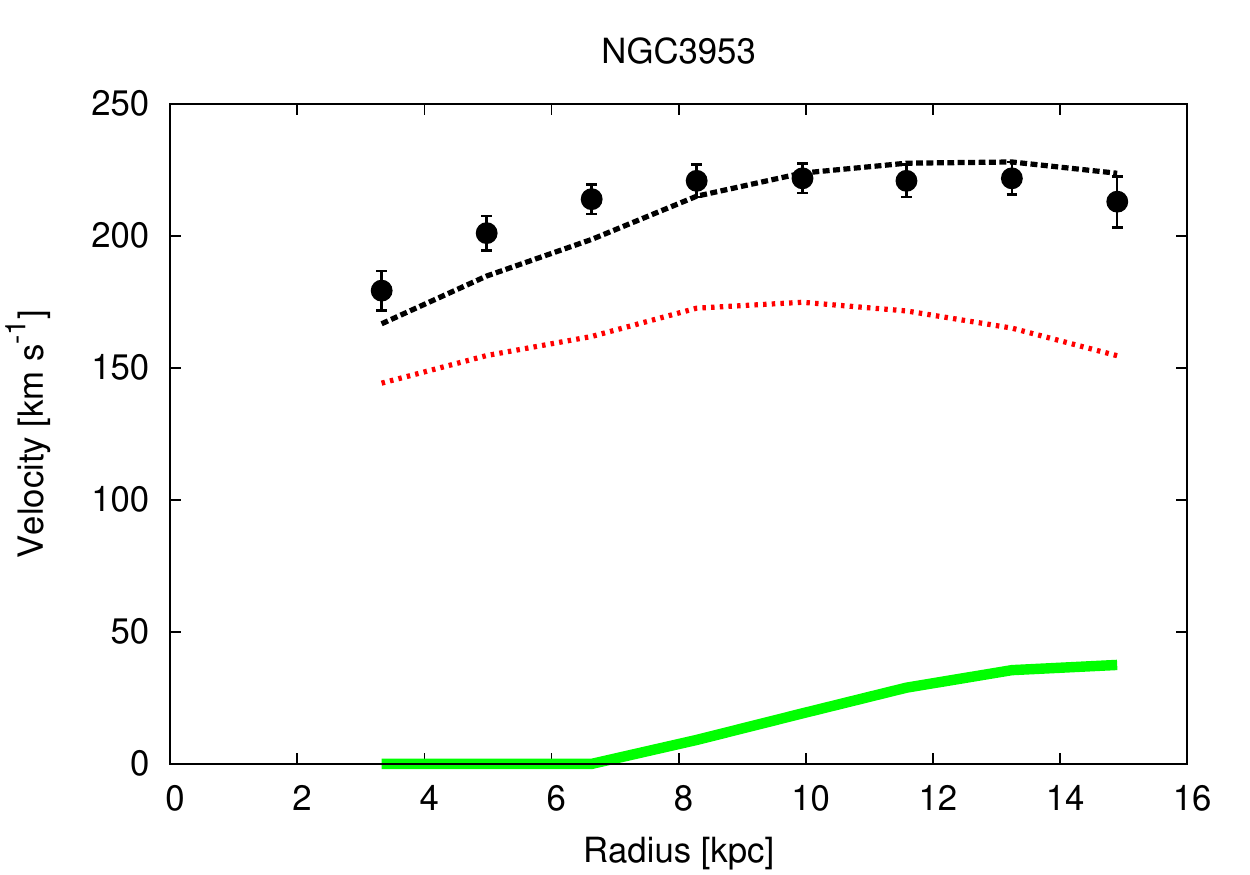} 
\caption{\bf The largest outlier in Table~\ref{tab:outliers}} {\small The MOND rotation curve for NGC~3953 (compare with the near-to-identical solution by Li et al.\cite{Li+18}) is shown as the black dotted curve and the observational data are plotted as black points with 1$\sigma$ uncertainties indicated by the horizontal bars (the contributions by gas and stars are indicated by the solid green and dotted red lines, respectively). This shows how the most discrepant galaxy (Table~\ref{tab:outliers}, $g_0 = 10^{-13.712}\,$km/s$^2$ $=0.16\,a_0$ with $M/L=1$ according to Rodrigues et al.\cite{Rodr18}) is well modelled by MOND (with $g_0=a_0$) if reasonable assumptions are made here to the distance $D=17.1\,$Mpc, stellar mass-to-light ratio in the disk, $M/L=0.5$, $i_{\rm best}=63.^{\circ}0$ yielding a reduced $\chi^2=3.6$ (compare with $D_{\rm obs}=16.0 \pm 1.8\,$Mpc, $M/L=0.59 \pm 0.1$ and $i_{\rm obs}=61.^{\circ}9\pm 1.^{\circ}0\,$ with a reduced $\chi^2=3.42$ in Li et al.\cite{Li+18}).  The analytical MOND model is clearly good despite a high reduced $\chi^2$, which may be due to a radial gradient in $M/L$ or other \emph{small} systematics.  A proper MOND representation of this galaxy would need to be a 3D model including a velocity ellipsoid at different radii and allow for a possible warp, a non-smooth and non-axially-symmetric matter distribution and the external field effect.
} \label{fig:outlier} \end{figure}

Of the 175 galaxies in the SPARC catalogue\cite{Lelli16a} used by Rodrigues et al.\cite{Rodr18}, there are 149 galaxies with quality flag $Q \leq 2$ and inclinations $i > 30^\circ$.  Amongst these are 59 (40\%) outliers whose inferred $\left| {\rm log}_{10} \left( \frac{g_{_0}}{a_{_0}} \right) \right | > 0.2$, being inconsistent with 0 at $>3\sigma$ confidence. By using only the most reliable rotation curves ($Q = 1$), this is reduced to 33/93 (35\%) outliers. If we exclude very nearly edge-on systems with $i > 85^\circ$ where dust obscuration can make it difficult to constrain the luminous matter distribution, then we are left with 26/78 (33\%) outliers.  Rodrigues et al.\cite{Rodr18} only vary distances, $D$, by $\pm 20\%$.  For galaxies where the $D$ estimates are based only on the Hubble law, the $D$ errors are underestimated. At the typical $D\approx 50\,$Mpc\cite{Li+18}, a Local Group-like peculiar velocity of 630~km/s\cite{Kogut93} would alone cause an error in $D$ of 19\%, being close to the range allowed by Rodrigues et al.\cite{Rodr18} Thus they only allow for $1\sigma$ variations in $D$. This suggests an outlier fraction of 32\%, close to the real fraction of claimed outliers.  Rejecting galaxies with no other type of $D$ measurement leaves just 5/29 (17\%) outliers (Table~\ref{tab:outliers}), challenging the analysis of Rodrigues et al.\cite{Rodr18}. Of these, figure A1 of Li et al.\cite{Li+18} shows that NGC 3953 (our Fig.~\ref{fig:outlier}) and 3992 are well fit with slight variations in the stellar mass-to-light ratio, $M/L$, while NGC 4183 is well fit if its distance is 28\% below the published value. Given the nominal 14\% distance uncertainty for this galaxy, this is plausible but falls outside the $\pm 20\,$\% range allowed by Rodrigues et al.\cite{Rodr18}.

The rotation curves of real galaxies are subject to a range of non-negligible effects: warps due to past encounters with other galaxies, differences in stellar populations arising from different star formation histories between galaxies and within different parts of the same galaxy, non-uniform mass distribution within a galaxy (Renzo's Rule\cite{FM12}) and the external field effect\cite{Haghi16} which changes the local Milgromian potential in this non-linear gravitational theory.

Figure~6 of Li et al.\cite{Li+18} explicitly shows that allowing $g_0$ to vary does not improve the MOND fits to the same galaxy rotation curves used by Rodrigues et al.\cite{Rodr18}. In view of the published work of Li et al.\cite{Li+18} which already demonstrated these galaxies to be well consistent with MOND, the conclusion reached by Rodrigues et al.\cite{Rodr18} cannot be upheld.  Instead, the results by Li et al.\cite{Li+18} support a simpler paradigm\cite{Disney08} where galaxies follow one gravitational law sourced only by the standard matter.

\clearpage

\clearpage

\clearpage

\begin{table}
  \centering
		\begin{tabular}{cccc}
			\hline
			Galaxy & $Log_{10} \left( \frac{g_{_0}}{a_{_0}}\right)$ & $\frac{\sigma_{_D}}{D}$ & Comments \\ [5pt]
			\hline
			NGC 3953 & $-0.88^{+0.57}_{-2.46}$ & 13.9\% & Consistent if $M/L$ varies slightly\\
			NGC 4183 & $-0.62^{+0.45}_{-0.16}$ & 13.9\% & Consistent if $D$ is 28\% below observed value\\
			UGC 07524 & $-0.51^{+0.41}_{-0.15}$ & 5.1\% & Irregular appearance, well fit with lower $i$ \cite{Swaters_2010} \\
			NGC 3992 & $-0.41^{+0.35}_{-0.26}$ & 9.7\% & Consistent if $M/L$ varies slightly \\
			NGC 2403 & $0.27^{+0.02}_{-0.03}$ & 5.1\% & Discrepant over very narrow range of radii \\
			\hline
		\end{tabular}
	\caption{ {\bf Properties of galaxies identified by Rodrigues et al.\cite{Rodr18} which are discrepant with MOND, showing their inferred $3\sigma$ uncertainty on $g_{_0}$.} This is an underestimate as they neglect inclination errors. Figure A1 of Li et al.\cite{Li+18} shows good overall consistency between predicted and observed rotation curves of these galaxies if some of the assumptions in Rodrigues et al.\cite{Rodr18} are relaxed slightly, for instance that $D$ is within 20\% of the observed value (NGC~4183) and that there are no radial gradients in $M/L$.}
  \label{tab:outliers}
\end{table}

\clearpage

\noindent{\bf References}
\bibliographystyle{naturemag}
\bibliography{/Users/pavel/PAPERS/BIBL_REFERENCES/kroupa_ref}

\clearpage

\noindent{\bf Acknowledgements}

\noindent Indranil Banik and Akram Hasani Zonoozi are  Alexander von Humboldt Fellows. Hosein Haghi is a DAAD visiting scholar. Behnam Javanmardi and J\"org Dabringhausen are grateful for the hospitality of the SPODYR group in Bonn and of the AIfA, where this work was done. Oliver M\"uller thanks the Swiss National Science Foundation for financial support. XW acknowledges support from the Natural Science Foundation of China (grant numbers 11503025 and 11421303), the Anhui Natural Science Foundation (grant number 1708085MA20), the ``Hundred Talents Project of Anhui Province'' and ``the Fundamental Research Funds for the Central Universities''. We thank the ``DAAD-Ostpartnerschaftsprogramm f\"ur 2018'' at the University of Bonn for funding exchange visits between Charles University in Prague and Bonn University.

\noindent{\bf Author contributions:}

\noindent Pavel Kroupa devised and wrote the manuscript and Indranil Banik devised the selection criteria.  Akram Hasani Zonoozi and Hosein Haghi calculated the rotation curve models.  Xufen Wu and Hongsheng Zhao contributed MOND expertise. Oliver M\"uller, Behnam Javanmardi and Joerg Dabringhausen contributed observational expertise on galaxies. All co-authors significantly contributed to the written record through comments and passages.

\noindent{\bf Statement on competing interests:}

\noindent The authors declare no competing interests.

\end{document}